\documentclass[conference]{IEEEtran}
\usepackage{graphicx}

\usepackage{cite}
\usepackage[cmex10]{amsmath}
\usepackage{threeparttable}
\usepackage{amsmath}
\usepackage{amssymb}
\usepackage{diagbox}

\usepackage{graphicx}
\usepackage{epstopdf}

\usepackage{algorithm}

\usepackage{algpseudocode}
\usepackage{float}

\def\BibTeX{{\rm B\kern-.05em{\sc i\kern-.025em b}\kern-.08em
    T\kern-.1667em\lower.7ex\hbox{E}\kern-.125emX}}

\begin{document}
\title{A Soft Cancellation Decoder for Parity-Check Polar Codes}
\author{\IEEEauthorblockN{Jiajie Tong, Huazi Zhang, Xianbin Wang, Shengchen Dai, Rong Li, Jun Wang}
\IEEEauthorblockA{Huawei Technologies Co. Ltd.\\
Email: \{tongjiajie, zhanghuazi, wangxianbin1, daishengchen, lirongone.li, justin.wangjun\}@huawei.com}}

\maketitle

\begin{abstract}
Polar codes has been selected as the channel coding scheme for 5G new radio (NR) control channel.
Specifically, a special type of parity-check polar (PC-Polar) codes was adopted in uplink control information (UCI).
In this paper, we propose a parity-check soft-cancellation (PC-SCAN) algorithm and its simplified version to decode PC-Polar codes.
The potential benefits are two-fold. First, PC-SCAN can provide soft output for PC-Polar codes, which is essential for advanced turbo receivers.
Second, the decoding performance is better than that of successive cancellation (SC). This is due to the fact that parity-check constraints can be exploited by PC-SCAN to enhance the reliability of other information bits over the iterations.
Moreover, we describe a cyclic-shift-register (CSR) based implementation ``CSR-SCAN'' to reduce both hardware cost and latency with minimum performance loss.
\end{abstract}


\IEEEpeerreviewmaketitle

\section{Introduction}
\subsection{Background and related works}
Polar codes, proposed by Arikan~\cite{Arikan}, has been selected as the channel coding scheme for 5G NR control channel. Polar codes with SC decoding is proved to achieve channel capacity at infinite length. To improve error-correction performance at short or moderate length, SC-list (SCL) decoding~\cite{Polar:SCL} is proposed by keeping $L$ codeword candidates during the sequential decoding process. In practice, cyclic redundancy check (CRC)~\cite{CA-Polar}, parity-check (PC)~\cite{PC-Polar} and convolutional code (CC)~\cite{PAC-Polar} bits can be inserted as outer codes to further improve the performance. They can be unified as pre-transformed polar codes~\cite{PT-Polar}.

Among the pre-transformed polar codes, PC-Polar codes exhibits low-complexity, flexibility and robust performance gain under fine-granularity evaluations~\cite{PC-Polar}. A simplified version of PC-Polar~\cite{Lisbon,Reno} is adopted by 3GPP\cite{NR} to enhance the performance of 5G UCI channel. The PC bit values are set to linear combinations of some preceding information bits, implemented by simple cyclic shift registers. Such a simple pre-transformation, like CA-Polar and PAC-Polar, can improve the code minimum distance~\cite{PT-Polar}. It is shown in~\cite{PC-Polar} that PC-Polar outperforms CA-Polar by $0.2\sim1.0$dB under SCL decoders with $L=8$.

\subsection{Motivation and Contribution}
The problem with SC and SCL decoders is they cannot output soft decoding results. The latter is essential for many applications such as decoding of concatenated codes~\cite{Turbo} and advanced turbo receivers~\cite{TurboRecv} (e.g., in multiple-input multiple-output and non-orthogonal multiple access). To address this issue, a soft-cancellation (SCAN) decoding~\cite{SCAN} is proposed to both provide soft output and improve bit error rate (BER) performance. If multiple iterations are allowed, the frame error rate (FER) performance can also improve over SC decoding. Unfortunately, current SCAN algorithms do not support PC-Polar codes, or any pre-transformed polar codes.

In this paper, we propose a parity-check soft-cancellation (PC-SCAN) algorithm and its simplified version to decode PC-Polar codes.
This is the \emph{first} attempt to implement a SCAN decoder on PC-polar codes.
Our algorithm incorporates an additional layer in factor graph to account for the PC constraints, and specifies a scheduling method to update the soft information.
Simulation results show better BER and FER (with more than one iterations).
An example of $N=64, K=32$ polar codes is given in Fig.~\ref{scanvssc}.

PC-SCAN requires an additional module on top of SCAN. The module updates the log-likelihood ratio (LLR) of the information/PC bits.
At the first glance, the updating rules are similar to those in LDPC codes. However, they turn out to be quite different due to the sequential scheduling of SCAN decoders.
Hence, we propose novel LLR updating rules to address this change. Furthermore, to facilitate hardware-friendly implementation, we propose a simplified version of PC-SCAN named CSR-SCAN. The simplification mainly exploits the regular structure of PC-Polar. We show that the performance loss is minimum ($<0.1$dB in FER) while the additional hardware area beyond the original SCAN decoder is very small.

\section{Tree Representation of Soft Cancellation}
\begin{figure}
\centering
\includegraphics[width=0.4\textwidth]{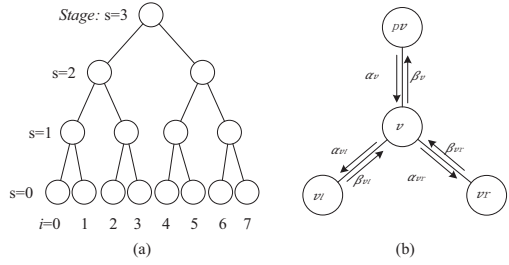} %
\caption{(a) decoding architecture as a binary tree; (b) node $\textit{v}$ received/response information}
\label{tree}
\end{figure}

In this section, we review the basic methods of SCAN to decode the basic Polar codes.
The SCAN decoder has a recursive architecture that is similar to the SC decoder.
Therefore, it can be represented as a binary tree traversal \cite{SSC}, as shown in Fig.~\ref{tree}(a).
Following the notations in \cite{SSC}, a node $v$ in a tree is directly connected to a parent node $p_{v}$, left child node $v_{l}$ and right child node $v_{r}$, respectively\footnote{A leaf node $v_{leaf}$ has no child node, and a root node $v_{root}$ has no parent node.}.
The stage of a node $v$ is defined by the number of edges between node $v$ and its nearest leaf node. All leaf nodes are at stage $s=0$.
The set of nodes of the subtree rooted at node $\textit{v}$ is denoted by $V_{v}$. Thus $V_{root}$ denotes the full binary decoding tree.
The set of all leaf nodes is denoted by $U$, the index of a leaf $\textit{u}$ \cite{SSC} is denoted by $l(u)$.

The set of all information bits are denoted by $\mathcal{I}$ and that of all frozen bits by $\mathcal{I}^{c}$.
We denote the information bit indices among the decoding tree.
\begin{equation*}
I = \left\{ l(u), u \in \mathcal{I} \right\}
\end{equation*}
Denote the frozen bit indices among the decoding tree by
\begin{equation*}
I^{c} = \left\{ l(u), u \in \mathcal{I}^{c} \right\}
\end{equation*}
Similar to \cite{SSC}, knowledge about decoded bits, e.g., in the form of log-likelihood ratio (LLR), is exchanged between neighboring nodes.
The root node $v_{root}$ receives LLRs $\alpha_{root}$ directly from the channel output.
As illustrated in Fig.~\ref{tree}(b), node $\textit{v}$ receives an LLR vector $\alpha_{v}$ from its parent $p_{v}$, and feeds back an LLR vector $\beta_{v}$ to its parent $p_{v}$.
The difference from \cite{SSC} is that here $\beta_{v}$ is no longer the hard decisions (partial sums), but soft information. It is called response LLR \cite{SCAN}.
If node $\textit{v}$ is at stage $\textit{s}$, vectors $\alpha_{v}$ and $\beta_{v}$ both have $2^{s}$ elements, respectively denoted by $\alpha_{v}^{i}$ and $\beta_{v}^{i}, i\in[0,2^{s}-1]$.

According \cite{SCAN}, a node $\textit{v}$ at stage $s$ passes information to its child nodes $\alpha_{v_{l}}$ and $\alpha_{v_{r}}$ as follows\footnote{In \cite{SCAN}, there is a bit-revisal step before forwarding each LLR vector. But this step is unnecessary \cite{systematic} and is removed in many other recent works.}:
\begin{equation}\label{alp_cal}
    \left\{
        \begin{array}{lr}
        \alpha_{v_{l}}^{i}=\alpha_{v}^{i}\boxplus(\beta_{v_{r}}^{i}+\alpha_{v}^{i+2^{s-1}}),i\in \left\{0,\cdots,2^{s-1}-1\right\} \\
        \alpha_{v_{r}}^{i}=(\alpha_{v}^{i}\boxplus\beta_{v_{l}}^{i})+ \alpha_{v}^{i+2^{s-1}},i\in \left\{0,\cdots,2^{s-1}-1\right\}
        \end{array}
    \right.
\end{equation}

The soft information feedback to its parent node is as follows
\begin{equation}\label{beta_cal}
\left\{
        \begin{array}{lr}
        \beta_{v}^{i} = \beta_{v_{l}}^{i}\boxplus(\alpha_{v}^{i+2^{s-1}} + \beta_{v_{r}}^{i}),i\in \left\{0,\cdots,2^{s-1}-1\right\} \\
        \beta_{v}^{i+2^{s-1}} = (\beta_{v_{l}}^{i}\boxplus\alpha_{v}^{i}) + \beta_{v_{r}}^{i},i\in \left\{0,\cdots,2^{s-1}-1\right\}
        \end{array}
    \right.
\end{equation}

Unlike SC, the SCAN decoder runs iteratively, which means binary tree can be traversed several rounds.
In the first iteration, we should initialize $\alpha_{root}$ as the channel output LLRs, and $\beta_{v},v\in V_{root}\setminus U_{root}$ as all zeros.
During the course of decoding, the vector $\beta_{v}$ will be updated by \eqref{beta_cal}. The algorithm terminates after a maximum of $T_{\max}$ iterations.

The soft messages $\beta_{v},v\in U_{root}$ at leaf nodes do not change over time. They are always either $+ \infty$ (for zero-valued frozen bits) or 0 (for information bits).

\section{Parity-Check Polar Codes}\label{sec:pc}
For PC-Polar codes, there are three types of bits, i.e., information bits, frozen bits and PC bits.
We address three PC-Polar schemes in this paper:
\begin{itemize}
\item NR(new radio)-PC-Polar as described in ``5G; NR; Multiplexing and channel coding''\cite{NR}, where there are three PC bits.
\item FC(full check)-PC-Polar as described in \cite{PC-Polar}, where all non-information bits are treated as PC bits. This gives the best performance.
\item MC(min-check)-PC-Polar as described in \cite{Lisbon}, where all non-information bits with specific row weight(s) are selected as PC bits. The row weight can be $w_{min}$, the minimum row weight in the information set, or $2\times w_{min}$, the second lowest row weight in the information set.
\end{itemize}

The only difference among NR-PC-Polar, FC-PC-Polar and MC-PC-Polar is the PC bits number/positions.
As said, NR-PC-Polar has a fixed number of PC bits. But for MC-PC-Polar, the PC bits number is controlled by a coefficient $\mathcal{A}$\footnote{The notation of the coefficient is originally $\alpha$ in \cite{PC-Polar}. To avoid conflict with the soft input $\alpha$ in this paper, we use $\mathcal{A}$ to rename the coefficient.}~\cite{PC-Polar}. The PC bit values are linear combinations of some preceding information bits. Such a pre-coding, or pre-transform, can improve the code minimum distance~\cite{PT-Polar}.

Algorithm~\ref{alg:PC-ENC} shows the encoding steps of a length-$N$ PC-Polar code, where $N=2^n$ is the mother code length. Comparing with the original Polar encoder, it only has an additional step 4.
\begin{algorithm}
\caption{Parity-check Polar encoding~\cite{PC-Polar}}
\begin{algorithmic}
\State {\bf(1)} Generate the information/frozen/PC bits positions.
\State {\bf(2)} Generate an all-zero sequence of length $N$.
\State {\bf(3)} Fill in the transmitted bit values at information positions, resulting in sequence $\bf s$.
\State {\bf(4)} PC pre-coding, get the new sequence $\bf q$.
\State {\bf(5)} Polar encoding, ${\bf q}\times {\bf G}$, where $\bf G$ is a $N\times N$ Kronecker matrix.
\end{algorithmic}
\label{alg:PC-ENC}
\end{algorithm}

In~\cite{PC-Polar}, a very simple way of PC pre-coding is proposed. The value of a PC bit $u \in \mathcal{P}$ is the sum of all its preceding information bits with a spacing of $L$ (or multiples of $L$). The PC constraints can be formulated as
\begin{equation}\label{equ:PC constraints}
{\bf s}[i] = \sum_{j<i\atop (i-j)\%L=0} {\bf s}[j],
\end{equation}
where $i$ is the index of the PC bit, and $j$ denotes the indices of all preceding information bits checked by the PC bit.

\begin{figure}
\centering
\includegraphics[width=0.45\textwidth]{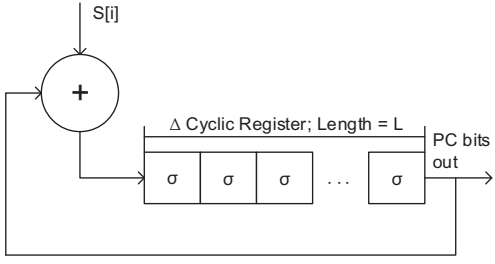} %
\caption{PC pre-coding circuit.}
\label{pre-coding}
\end{figure}

The PC pre-coding can be easily implemented by cyclic shift registers $\sigma$ of length $L$. The circuit is very simple, as shown in Fig.~\ref{pre-coding}, where a register state is updated by
\begin{equation}\label{equ:pre-coding}
\sigma_{x} = \sigma_{x} \oplus {\bf s}[i],\; x = i\%L,\; i < N.
\end{equation}

Traditionally, its decoder is modified from SC or SCL by adding a PC bit calculation module. During path extension and pruning, only the paths that satisfy the PC constraints are kept. The process resembles hard decoding where the PC bit values serve as ``hard constraints'' to select the valid paths.

\section{Parity Check Soft Cancellation}\label{sec:nodes}
In this section, we introduce a novel soft cancellation decoding algorithm, namely PC-SCAN, that supports PC-Polar codes\footnote{In principle, the method also supports any pre-transformed polar codes~\cite{PT-Polar}, while our focus is on PC-Polar codes~\cite{PC-Polar} due to its simplicity and performance.}. Unlike SC/SCL decoders, the proposed method treats PC bits as ``soft constraints'' to process the soft information generated by SCAN. Consequently, PC-SCAN can generate the soft outputs.

We still use $\mathcal{I}$ and $\mathcal{F}$ to denote the set of information bits and frozen bits, respectively. The set of PC bits is denoted by $\mathcal{P}$.
The frozen and PC bit indices among the decoding tree's leaf nodes $U$ are respectively denoted by
\begin{align*}
F = \left\{ l(u),u \in \mathcal{F} \right\}, \\
P = \left\{ l(u),u \in \mathcal{P} \right\}.
\end{align*}
By definition, $\mathcal{F}\cup \mathcal{P}= \mathcal{I}^{c}$ and $F_{v}\cup P_{v}=I_{v}^{c}$.

According to \eqref{equ:PC constraints}, only bit positions with a spacing of $L$ (or multiples of $L$) will engage in the same PC constraint.
Based on this observation, we can separate the leaf nodes in $\mathcal{I}$ and $\mathcal{P}$ into $L$ independent groups.
Specifically, the $i$-th information/PC bit belongs to the $i\%L$-th group, where the group index takes value in $\{0,1,\cdots,L-1\}$.
Since the nodes in the different groups are independent in terms of PC constraints, we call each group a ``parity-check chain''.

To facilitate description, we specify the following notations.
Denote by $\mathcal{I}(u)$ the set of all information bits checked by a PC bits $u \in \mathcal{P}$.
The bit indices in a ``checked information set'' $\mathcal{I}(u), u \in \mathcal{P}$ is denoted by
\begin{equation*}
I(u) = \left\{ l(u'),u' \in \mathcal{I}(u) \right\}, \ \text{for} \ u \in \mathcal{P}.
\end{equation*}
The union set of all \emph{checked} information bits is denoted by
\begin{equation}\label{equ:checked_info_set}
\mathcal{\tilde{I}} = \bigcup_{u\in\mathcal{P}}\mathcal{I}(u).
\end{equation}
The bit indices in $\mathcal{\tilde{I}}$ are denoted by
\begin{equation*}
\tilde{I} = \left\{ l(u),u \in \mathcal{\tilde{I}} \right\}.
\end{equation*}

The remaining set of all \emph{unchecked} information bits is denoted by
\begin{equation*}
\mathcal{\tilde{I}}^{c} = \mathcal{I} \setminus \mathcal{\tilde{I}}
\end{equation*}
The bit indices in $\mathcal{\tilde{I}}^{c}$ are denoted by
\begin{equation*}
\tilde{I}^{c} = \left\{ l(u),u \in \mathcal{\tilde{I}}^{c} \right\}.
\end{equation*}

Similarly, denote by $\mathcal{P}(u)$ the set of all PC bits \emph{checking} an information bits $u \in \mathcal{\tilde{I}}$.
The bit indices in a ``checking PC set'' $\mathcal{P}(u), u \in \mathcal{\tilde{I}}$ is denoted by
\begin{equation*}
P(u) = \left\{ l(u'),u' \in \mathcal{P}(u) \right\}, \ \text{for} \ u \in \mathcal{\tilde{I}}.
\end{equation*}

The PC-SCAN algorithm first constructs a binary-tree in the recursive manner as in the original SCAN decoder, and then traverse the tree. It can be divided into two parts:
\begin{itemize}
\item Traverse the tree and apply \eqref{alp_cal} and \eqref{beta_cal} to calculate the input/output soft information of each node $\textit{v}$ when entering/leaving the subtree $V_{v}$. When the tree traversal reaches a leaf node $u$, such that $u\notin\{\mathcal{P}\,\cup \,\mathcal{\tilde{I}}\}$, we directly obtain the soft feedback $\beta$ as follows
\begin{equation}\label{beta_function_org0}
\beta_{u}=\infty, \ \text{for} \ u \in \mathcal{F}.
\end{equation}
\begin{equation}\label{beta_function_org2}
 \beta_{u}=0, \ \text{for} \ u \in \mathcal{\tilde{I}}^{c}.
\end{equation}
\item The second part calculates the soft feedback $\beta$ at leaf nodes such that $u\in \{\mathcal{P}\,\cup \,\mathcal{\tilde{I}}\}$. These are the information/PC bits that engage in PC constraints. The $L$ ``parity-check chains'' are modelled by $L$ tanner graphs representing the parity-check relationships between information bits and PC bits. As will be described shortly, we use leaf nodes input $\alpha_{u}$ to calculate the soft feedback $\beta_{u}$.
\end{itemize}

\begin{figure*}
\centering
\includegraphics[width=0.98\textwidth]{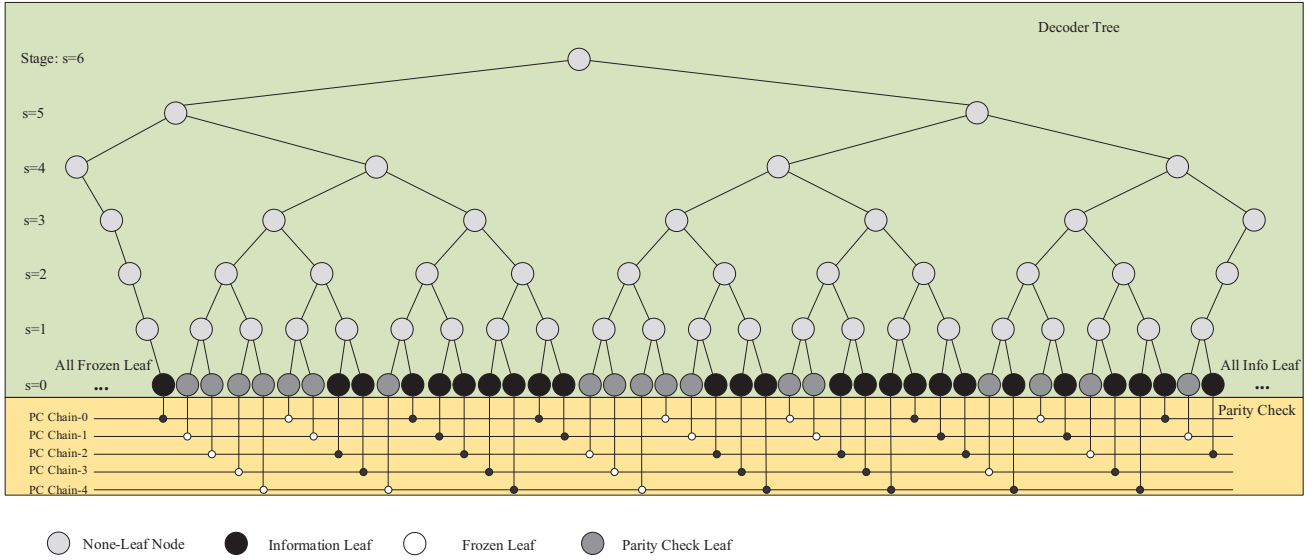} %
\caption{Connections between SC decoding tree and parity check constraints}
\label{pclevel}
\end{figure*}

Fig.~\ref{pclevel} shows an example of the connections between the original SCAN decoding tree and parity check chains. In this example, the number of information bits (including CRC bits) is 32 and the code length is 64. Specifically, the reliability order is determined by polarization weight (PW)~\cite{PW}, and the number and positions of PC bits are determined as in FC-PC-Polar~\cite{PC-Polar}. In the interest of space, the leaf nodes before the first information bit and after the last parity check bit are omitted, as they do not affect parity-check decoding. The length of cyclic shift registers is $L=5$, resulting in five independent parity-check chains.

The set of bits in parity-check chain-0 is denoted by
\begin{equation*}
\mathcal{U}(0) = \left\{u, u \in U \ \text{and} \ l(u)\%5=0 \right\}
\end{equation*}
According the Fig.~\ref{pclevel}, the indices of $u\in\mathcal{U}(0)$ are grouped as follows
\begin{equation*}
\left\{
        \begin{array}{lr}
        \{0,5,10\} \subset F \\
        \{20,35,40,50\} \subset P \\
        \{15,25,30,45\} \subset \tilde{I}\\
        \{55,60\} \subset \tilde{I}^{c}\\
        \end{array}
    \right.
\end{equation*}

\begin{figure}
\centering
\includegraphics[width=0.45\textwidth]{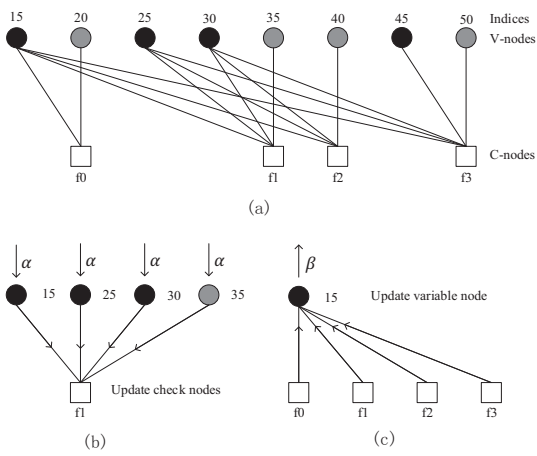} %
\caption{(a)\, The tanner graph of PC Chain-0; (b)\, Check nodes update (input $\alpha$);
(c)\, Cariable node update (output $\beta$).}
\label{tanner}
\end{figure}
The PC constraints are illustrated by a tanner graph is plotted in Fig.~\ref{tanner}(a). It includes all the checked information bits and PC bits in the PC chain-0 and model their parity-check relationship. All the checked information bits and PC bits are regarded as variable nodes in the tanner graph. The number of check nodes is equal to the number of PC bits in this PC-chain, as each check node represents a PC constraint. In this example, there are eight variable nodes and four check nodes.

Fig.~\ref{tanner}(b) and Fig.~\ref{tanner}(c) show how to update the variable nodes' soft output $\beta$ with the soft input $\alpha$ from the SCAN decoder.

For a PC bit $u\in \mathcal{P}$, its soft output $\beta$ is calculated by
\begin{align}\label{beta_function_org1}
\beta_{u}=\lambda_{\mathcal{P}} \times &(\alpha_{u'_{0}}\boxplus\alpha_{u'_{1}}\boxplus\cdots\boxplus\alpha_{u'_{N}}), \\ \nonumber
& \text{where} \ \{u'_{0},\cdots,u'_{N}\} = \mathcal{I}(u) \ \text{for} \ u \in \mathcal{P}.
\end{align}

For an information bit $u \in \mathcal{\tilde{I}}$, its soft output $\beta$ is
\begin{equation}\label{beta_function_org3}
\beta_{u}=\sum_{u' \in \mathcal{P}(u)}\eta_{u'}, \ \text{for} \ u \in \mathcal{\tilde{I}},
\end{equation}
where $\eta_{u'}$ is contributed by its neighboring PC bit $u'$
\begin{align*}
\eta_{u'} = \lambda_{\mathcal{\tilde{I}}} \times &(\alpha_{u'}\boxplus\alpha_{u''_{0}}\boxplus\cdots\alpha_{u''_{N}}),\\
& \ \text{where} \ \{u''_{0},\cdots,u''_{N}\} = \mathcal{I}(u') \setminus {u} \ \text{for} \ u' \in \mathcal{P}(u).
\end{align*}
Note that $\lambda_{\mathcal{P}}$ and $\lambda_{\mathcal{\tilde{I}}}$ are non-negative damping factors, which vary over code lengths and iterations. They are usually determined empirically.

In practice, the ``box-plus'' operation $\boxplus$ is implemented in an approximate but hardware-friendly way
\begin{align}\label{f-function}
x \boxplus y & \approx f(x,y) = \\ \nonumber
& (1-2({sign(x)\oplus sign(y)})) \times min(abs(x),abs(y)),
\end{align}
where $sign(x)$ is the sign of $x$, $abs(x)$ is the absolute value of $x$, and $min$ is minimum of input variables.

The $f$-function can also take multiple inputs
\begin{align*}
x_{1}\boxplus\cdots\boxplus & x_{n} \approx f(x_{1},\cdots,x_{n}) = \\
&(1-2({sign(x_{1})\oplus \cdots \oplus sign(x_{n})})) \\
& \times min(abs(x_{1}),\cdots,abs(x_{n})).
\end{align*}

With the above simplification, \eqref{beta_function_org1} can be rewritten as
\begin{equation}\label{beta_function_org1_opt}
\beta_{u}=\lambda_{\mathcal{P}}\times f(\alpha_{u'_{0}},\alpha_{u'_{1}},...,\alpha_{u'_{N}})\mid_{\{u'_{0}...u'_{N}\} = \mathcal{I}(u)}
\end{equation}

Likewise, \eqref{beta_function_org3} can be rewritten as
\begin{equation}\label{beta_function_org3_opt}
        \beta_{u}=\sum_{u' \in \mathcal{P}(u)}\lambda_{\mathcal{\tilde{I}}}\times f(\alpha_{u'},\alpha_{u''_{0}},...,\alpha_{u''_{N}})\mid_{\{u''_{0},...u''_{N}\} = \mathcal{I}(u') \setminus {u}}
\end{equation}

PC-SCAN follows a special scheduling that sequentially updates the soft information of leaf nodes. Specifically, the soft input $\alpha$ from SCAN are obtained successively over time. When updating the soft output $\beta$ of an information/PC bit, the associated soft input $\alpha$ from succeeding bits are not available. This is in contrast to belief propagation (BP) algorithms which can simultaneously update all leaf nodes.

For PC-SCAN, a detailed description of scheduling and node updating operations is described in Algorithm~\ref{alg:PC-SCAN}.

\begin{algorithm}
\caption{Parity-check Soft Cancellation (PC-SCAN)}
\begin{algorithmic}
\State {\bf Input: $\mathcal{I},\mathcal{F},\mathcal{P},\alpha_{root},T_{\max}$}
\State {\bf Output: $\beta$}
\State Derive $\mathcal{\tilde{I}}$ according to \eqref{equ:checked_info_set};
\State {\bf soft\_cancellation\_iterations}
\State \qquad for ($t=0$;\,$t<T_{\max}$;\,$t=t+1$)
\State \qquad\quad Call {\bf recursive\_decoder($v_{root},\alpha_{root}$)};
\State {\bf recursive\_decoder($v,\alpha_v$)}
\State \qquad Calculate $\alpha_{v_l},\alpha_{v_r}$ according to \eqref{alp_cal};
\State \qquad if ($v\notin U$)
\State \qquad \quad Call {\bf recursive\_decoder($v_l,\alpha_{v_l}$)};
\State \qquad \quad Call {\bf recursive\_decoder($v_r,\alpha_{v_r}$)};
\State \qquad else
\State \qquad \quad Call {\bf leaf\_decoder($u_l,\alpha_{u_l}$)};
\State \qquad \quad Call {\bf leaf\_decoder($u_r,\alpha_{u_r}$)};
\State \qquad Calculate $\beta_v$ according to \eqref{beta_cal};
\State {\bf leaf\_decoder($u,\alpha_u$)}
\State \qquad if ($u\in\mathcal{P}$)
\State \qquad \quad Calculate $\beta_u$ according to \eqref{beta_function_org1_opt};
\State \qquad else if ($u\in\mathcal{\tilde{I}}$)
\State \qquad \quad Calculate $\beta_u$ according to \eqref{beta_function_org3_opt};
\State \qquad else if ($u\in\mathcal{F}$)
\State \qquad \quad $\beta_u=\infty$;
\State \qquad else
\State \qquad \quad $\beta_u=0$;
\end{algorithmic}
\label{alg:PC-SCAN}
\end{algorithm}

\begin{figure}
\centering
\includegraphics[width=0.45\textwidth]{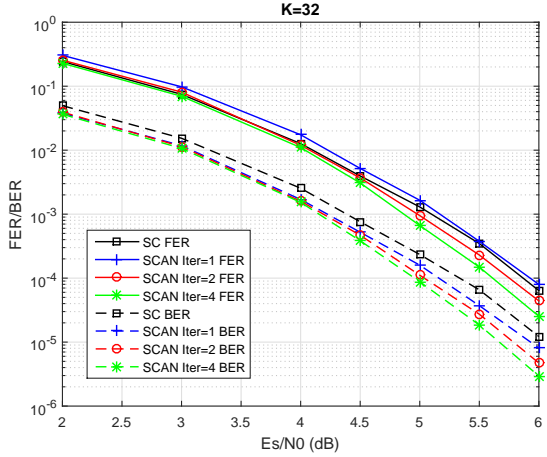} %
\caption{Performance of FC-PC-Polar with $N=64,K=32,\mathcal{A}=0.5$.}
\label{scanvssc}
\end{figure}

The FER and BER performances under PC-SCAN is shown in Fig.~\ref{scanvssc}, in which the code construction is given in Fig.~\ref{pclevel}.
In this case, the FER of PC-SCAN is similar to that of SC after the first iteration, but the BER is better by 0.1dB. After a few iterations, the PC-SCAN decoder outperforms SC on both FER and BER.
Besides, PC-SCAN can output soft decoding results.

\section{Cyclic-shift-register-based Hardware}
The leaf node decoder characterized by \eqref{beta_function_org0}, \eqref{beta_function_org2}, \eqref{beta_function_org1_opt} and \eqref{beta_function_org3_opt} can be further simplified to facilitate hardware implementation. While \eqref{beta_function_org1_opt} and \eqref{beta_function_org3_opt} support the soft decoding of any pre-transformed polar codes~\cite{PT-Polar} (e.g., CA-Polar~\cite{CA-Polar}, and PAC-Polar~\cite{PAC-Polar}), they do not exploit the regular structure of PC-Polar~\cite{PC-Polar} for further simplification.

We propose a cyclic-shift-register (CSR) based soft feedback generator. Compared with the general method of \eqref{beta_function_org1_opt} and \eqref{beta_function_org3_opt}, it does not need to store the factor graph in Fig.~\ref{tanner} and all the variables. This is a huge complexity reduction given the irregularity of the graph and the number of variables therein.

The LLR calculation can also be simplified. For information bits, we found that the soft feedback can be set to zero. For PC bits, the damping factor $\lambda_{\mathcal{P}}$ can be set to one to avoid multiplication operations. As will be shown, both simplifications incur only negligible performance loss.

The simplified soft feedback generation is
\begin{equation}\label{beta_function_all_opt2}
\left\{
        \begin{array}{lr}
        \beta_{u}=\infty,  \ \text{for} \ u \in \mathcal{F}\\
        \beta_{u}= f(\alpha_{u'_{0}},\alpha_{u'_{1}},...,\alpha_{u'_{N}})\mid_{\{u'_{0}...u'_{N}\} = \mathcal{I}(u)}, \ \text{for} \ u \in \mathcal{P}\\
        \beta_{u}=0,  \ \text{for} \ u \in \mathcal{I}
        \end{array}
    \right.
\end{equation}

\begin{figure}[H]
\centering
\includegraphics[width=0.49\textwidth]{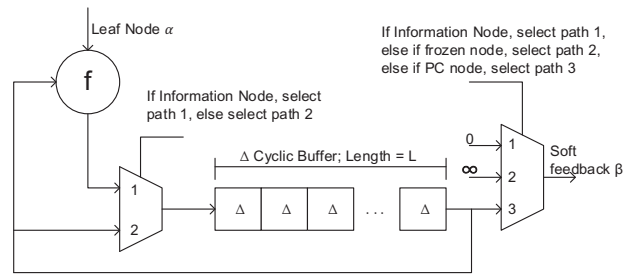} %
\caption{Soft feedback generator for PC-SCAN leaf decoder}
\label{pc-cyclic}
\end{figure}

The simplified circuit is shown in Fig.~\ref{pc-cyclic}, and described as follows.
We set up an array of registers ``$\Delta$'' of length $L$. Each register processes one parity-check chain.
The registers only store the intermediate states that are necessary for generating the next soft feedback.

Before each iteration, all register states are reset to zero.
As the registers shift, the $f$-function in \eqref{beta_function_all_opt2} can be calculated on the fly.
Upon decoding an information bit $u \in \mathcal{I}$, the input soft LLR $\alpha_{u}$ is accumulated into the register $\Delta$ as follows
\begin{equation}
\Delta_{x} = f(\Delta_{x}, \alpha_{u}),  \ \text{for} \ u\in\mathcal{I},~x=l(u)\% L.
\end{equation}

Upon decoding a PC bit $u$, the soft feedback $\beta_{u}$ can be immediately obtained as
\begin{equation}
\beta_{u} = \Delta_{x},  \ \text{for} \ u\in\mathcal{P},~x=l(u)\% L.
\end{equation}

\begin{figure*}
\centering
\includegraphics[width=0.75\textwidth]{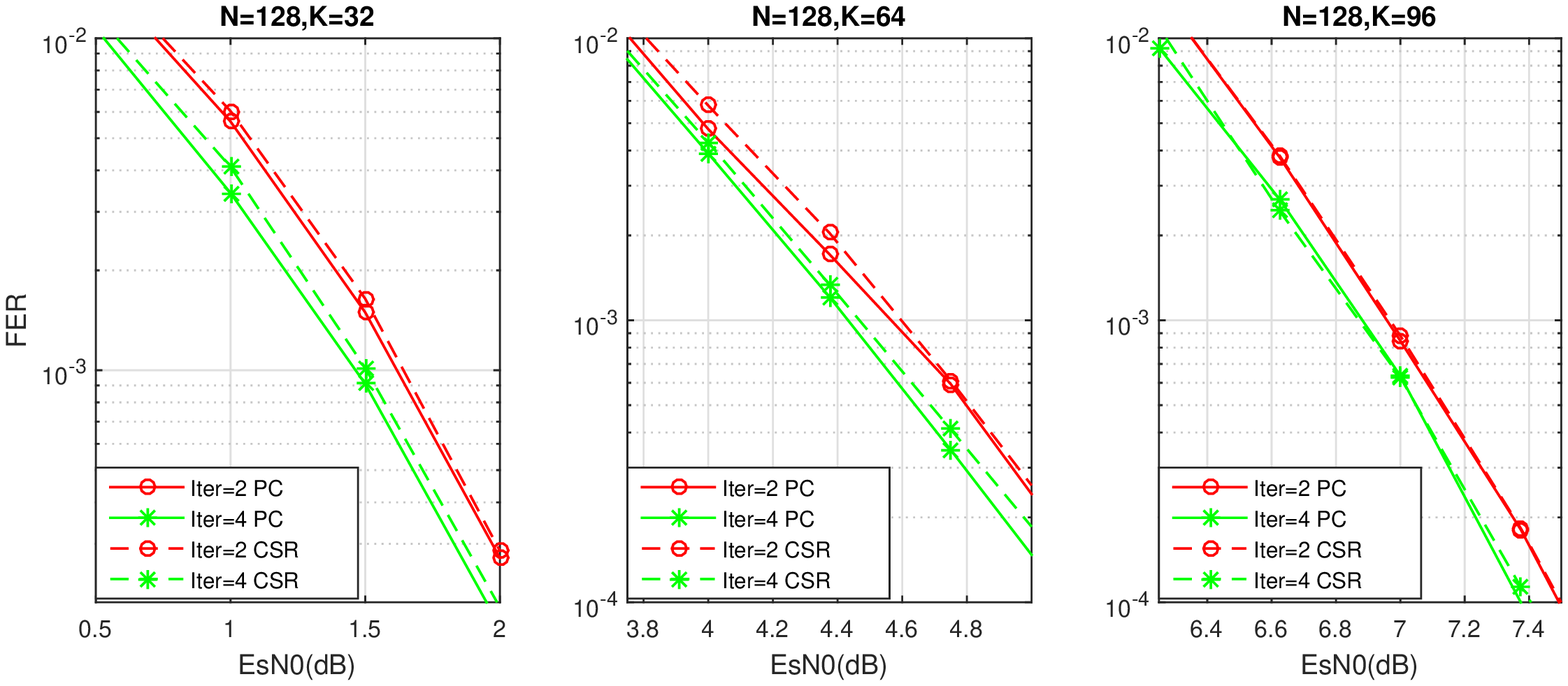} %
\caption{FC-PC-Polar, N=128, K=[32,64,96], $\mathcal{A}$=1.0, performances of PC-SCAN using the damping factor $\lambda_{\mathcal{\tilde{I}},best}$=[0.67,0.67,0.5] and CSR-SCAN.}
\label{zerocoef128}
\end{figure*}
\begin{figure*}
\centering
\includegraphics[width=0.75\textwidth]{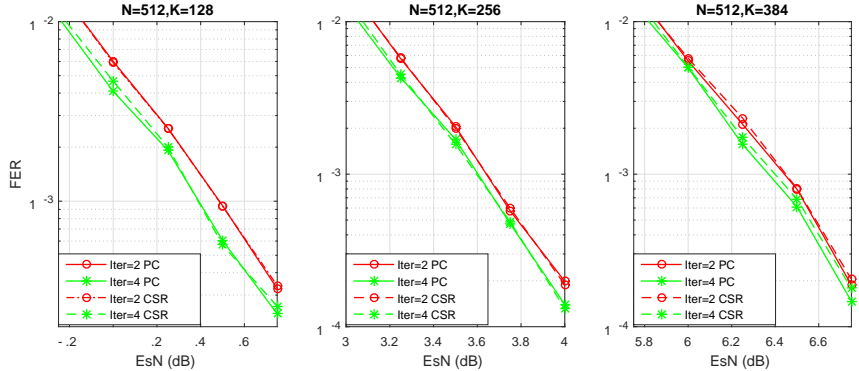} %
 \caption{MC-PC-Polar, N=512, K=[128,256,384], $\mathcal{A}$=1.5, performances of PC-SCAN using the damping factor $\lambda_{\mathcal{\tilde{I}},best}$=[0.67,0.67,0.5] and CSR-SCAN.}
\label{zerocoef512}
\end{figure*}
\begin{figure*}
\centering
\includegraphics[width=0.75\textwidth]{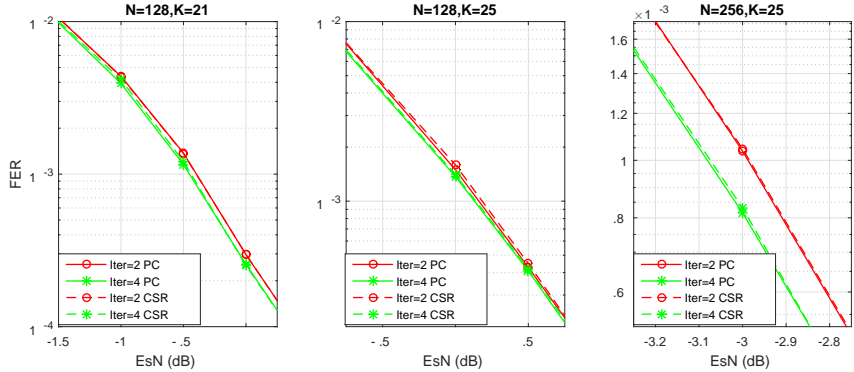} %
 \caption{NR-PC-Polar, N=128, K=[21,25] and N=256, K=25, performances between PC-SCAN using the damping factor $\lambda_{\mathcal{\tilde{I}},best}$=[0.67,0.5,0.5] and CSR-SCAN.}
\label{zerocoefNR}
\end{figure*}

Therefore, to implement PC-SCAN, we only need a small additional circuit on top of SCAN, thanks to the structural regularity of PC-Polar~\cite{PC-Polar}. We name this optimized method  CSR-SCAN decoder.
The circuit comprises only an $f$-function, defined in \eqref{f-function}, and a cyclic buffer of length $L$. As shown in Fig.~\ref{pc-cyclic}, it is very similar to the PC pre-coding circuit in Fig.~\ref{pre-coding}. Both are very hardware friendly and highly efficient in terms of area, power and latency.

\section{Performances}
\begin{figure*}
\centering
\includegraphics[width=0.75\textwidth]{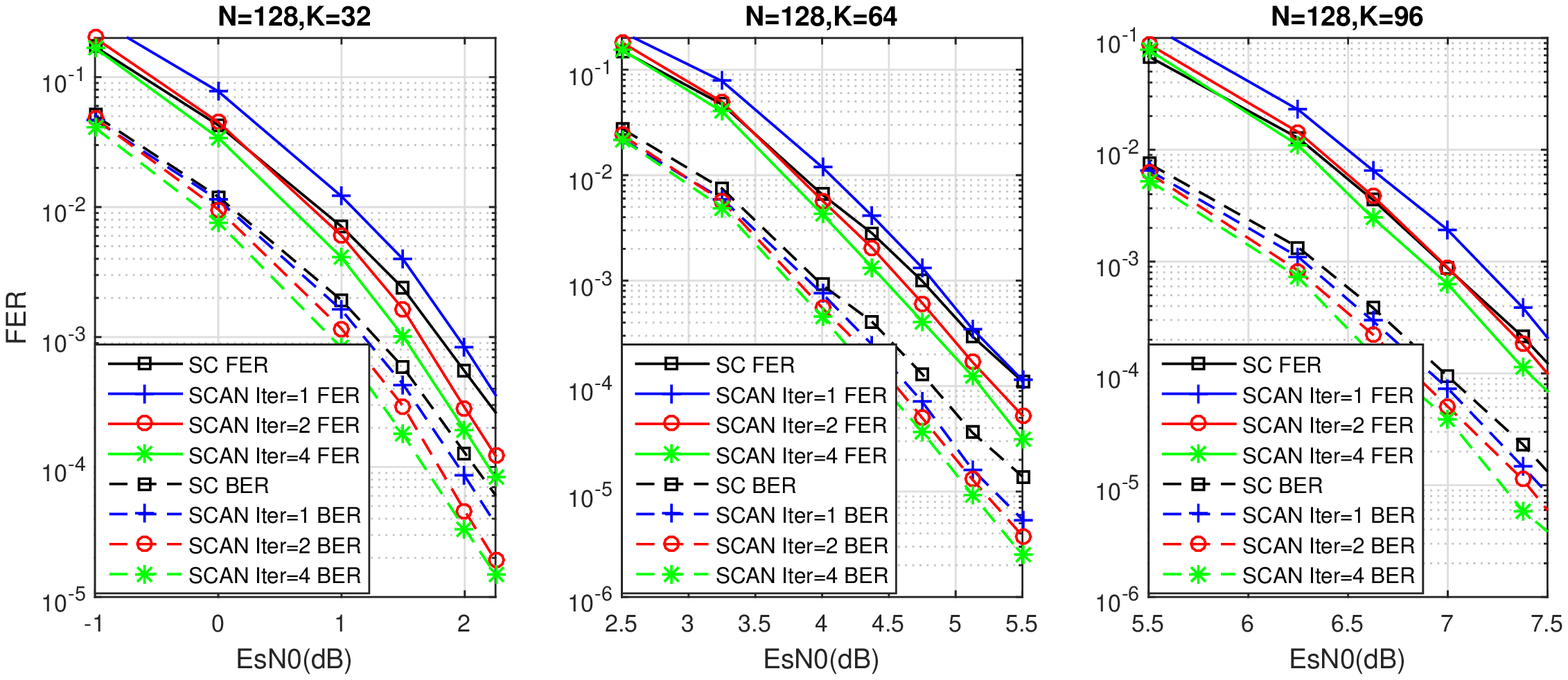} %
\caption{FC-PC-Polar, N=128, K=[32,64,96], $\mathcal{A}$=1.0, performances of SC decoder and CSR-SCAN decoder.}
\label{comp128fc}
\end{figure*}
\begin{figure*}
\centering
\includegraphics[width=0.75\textwidth]{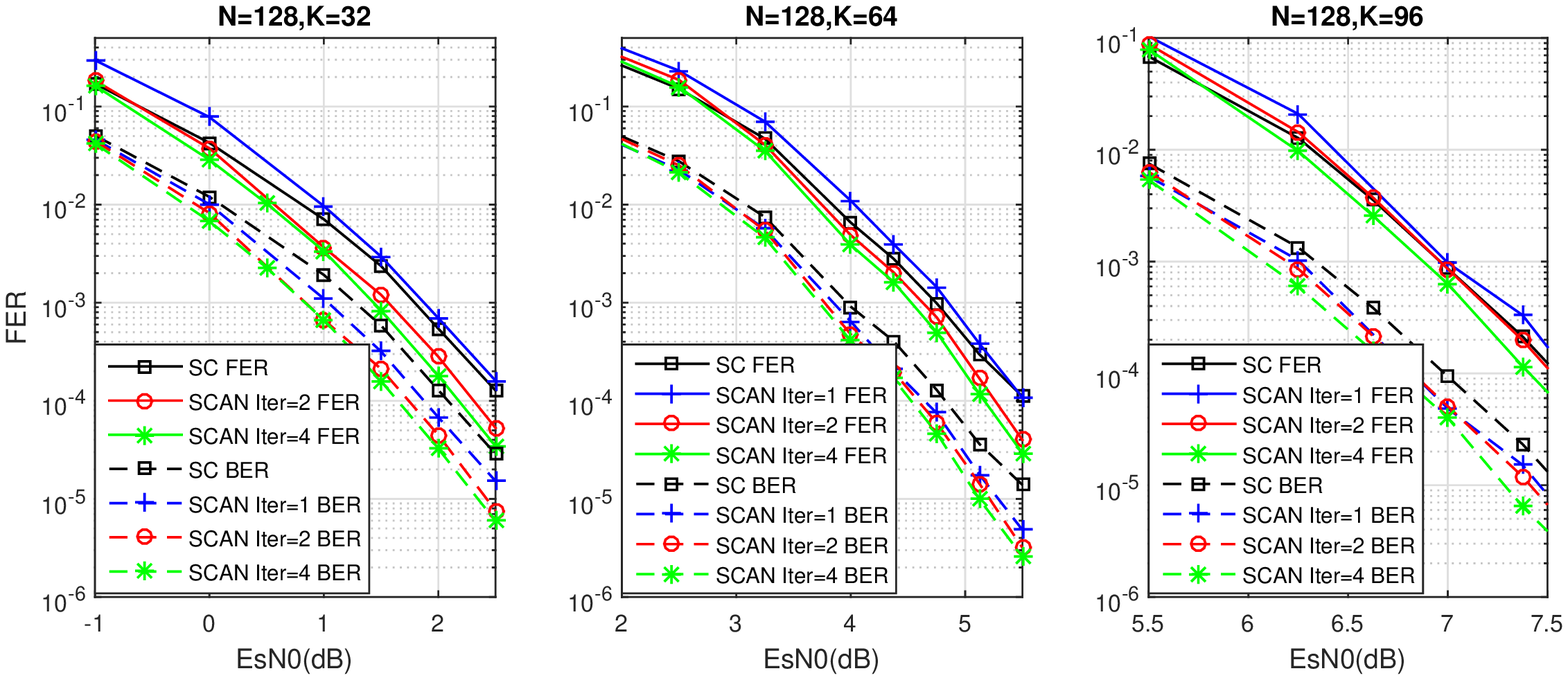} %
\caption{MC-PC-Polar, N=128, K=[32,64,96], $\mathcal{A}$=1.0, performances of SC decoder and CSR-SCAN decoder.}
\label{comp128mc}
\end{figure*}
\begin{figure*}
\centering
\includegraphics[width=0.75\textwidth]{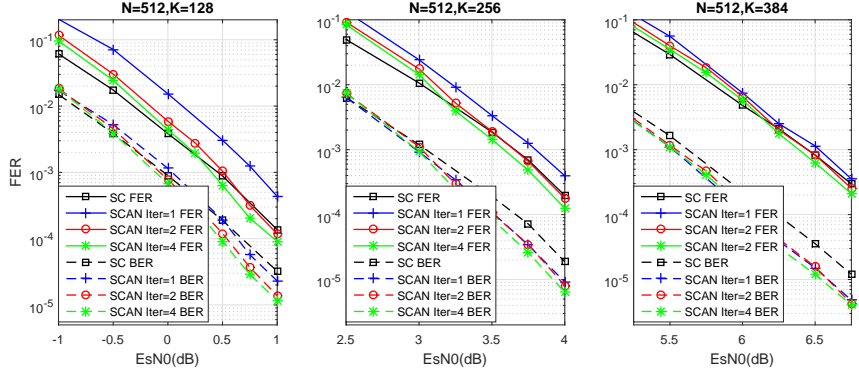} %
\caption{FC-PC-Polar, N=512, K=[128,256,384], $\mathcal{A}$=1.5, performances of SC decoder and CSR-SCAN decoder.}
\label{comp512fc}
\end{figure*}
\begin{figure*}
\centering
\includegraphics[width=0.75\textwidth]{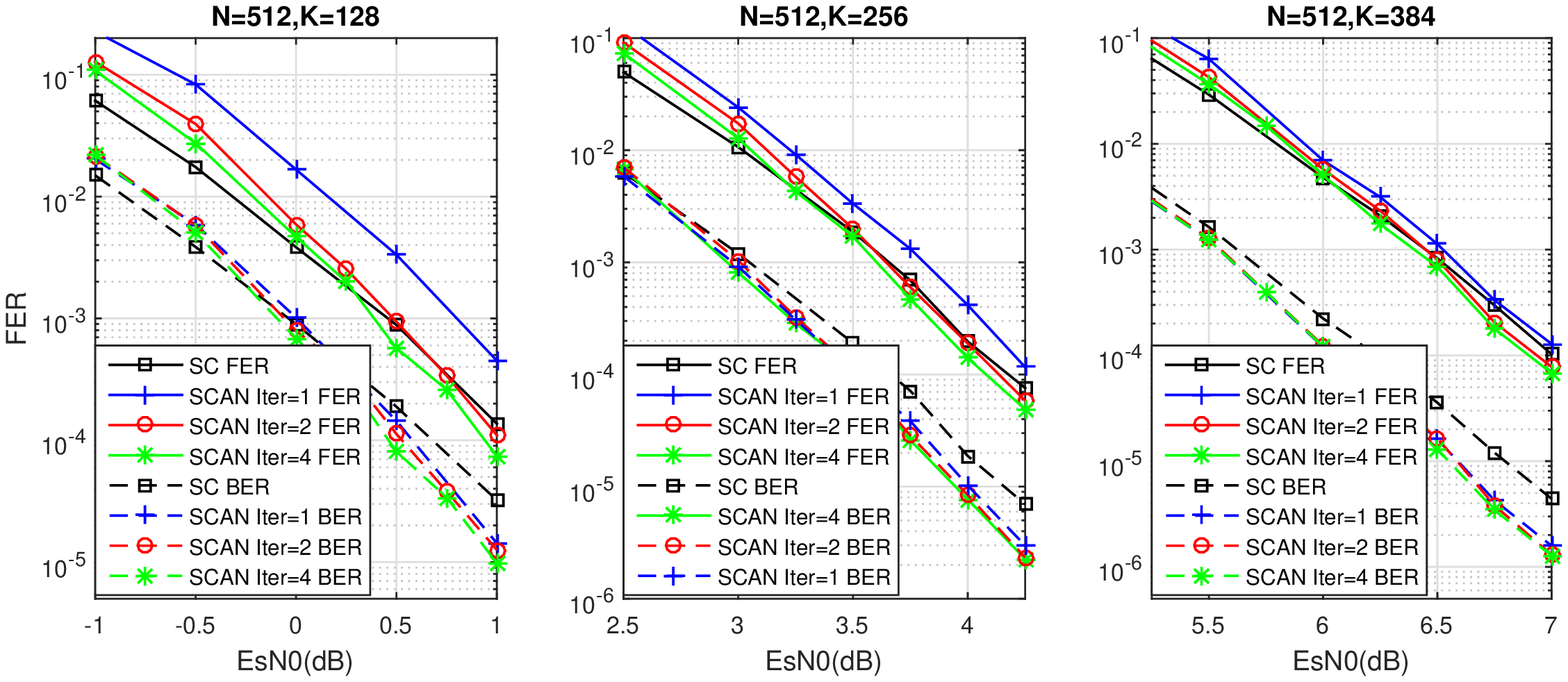} %
\caption{MC-PC-Polar, N=512, K=[128,256,384], $\mathcal{A}$=1.5, performances of SC decoder and CSR-SCAN decoder.}
\label{comp512mc}
\end{figure*}
\begin{figure*}
\centering
\includegraphics[width=0.75\textwidth]{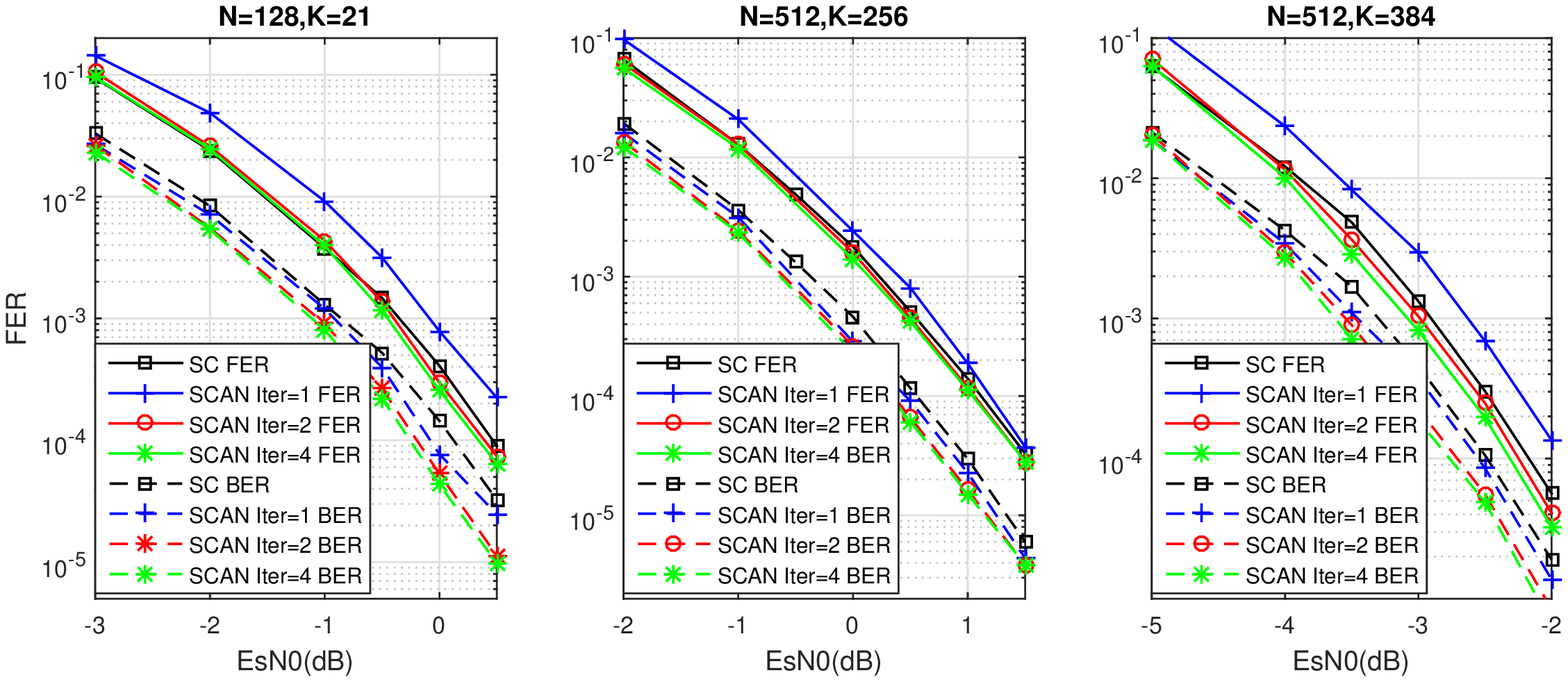} %
 \caption{NR-PC-Polar, N=128, K=[21,25] and N=256, K=25, performances of SC decoder and CSR-SCAN decoder.}
\label{compNR}
\end{figure*}

We simulate the performances of PC-SCAN decoder using the best damping factor $\lambda_{\mathcal{\tilde{I}}}$\footnote{The damping factor is obtained empirically through simulations.} and CSR-SCAN decoder within all the PC-types introduced in section~\ref{sec:pc}. The results in Fig.~\ref{zerocoef128}, Fig.~\ref{zerocoef512} and Fig.~\ref{zerocoefNR} show such simplifications incur minimum performance loss.
After the first iteration, the optimized circuit produces the same performance as the original one. With more iterations, a little performance loss can be observed.
For all simulated cases, the FER loss is within 0.05dB.

We also compare the performance between SC decoder and CSR-SCAN decoder. We show the FER and BER performance comparison in the Fig.~\ref{comp128fc} $\sim$ Fig.~\ref{compNR}. We find that in the cases of shorter code length ($N$=128), the FER of CSR-SCAN after the first iteration is 0.1dB inferior to that of SC decoder at $FER=10^{-3}$, while the BER is 0.1dB better. After the second iteration, both FER and BER of CSR-SCAN are better than those of SC decoder. In the cases of long codes ($N$=512), CSR-SCAN's slope at the waterfall region is better than that of SC decoder, and the performances after the second iteration are similar to that of SC decoder when $FER=10^{-3}$. The NR-PC-Polar has 3 PC-bits, and exhibits less gain from the 2-rd to the 4-th iteration. The FER of CSR-SCAN's second iteration is better than that of SC-decoder by 0.1dB.

\section{Conclusion}
In this paper, we propose a soft-output PC-SCAN decoder for pre-transformed polar codes including CA-Polar, PC-Polar and PAC-Polar.
The proposed algorithm treats PC bits as ``soft constraints'' to process the soft information generated by SCAN.
An additional factor graph is formulated to process these constraints.
According to the factor graph, four types of leaf nodes are defined, i.e., frozen/PC/checked information/unchecked information, and decoded accordingly.
With focus on the hardware implementation, a cyclic-shift-register-based optimization is proposed to dramatically reduce the hardware overhead named CSR-SCAN decoder.
Finally, performance simulations show both the gain from PC bits and the insignificant performance loss due to CSR-SCAN decoder.

\end{document}